\documentclass[twocolumn,english,aps,prl,groupedaddress]{revtex4}
\usepackage{subfigure}
\usepackage{graphics}

\usepackage{graphicx}
\usepackage[T1]{fontenc}
\usepackage[latin9]{inputenc}
\usepackage{color}
\usepackage{ifsym}
\usepackage{wasysym}
\usepackage{hyperref}
\usepackage{babel}
\usepackage[normalem]{ulem}
\usepackage{url}

\begin{document}

\title{Elastic wake instabilities in a creeping flow between two obstacles}
\author{Atul Varshney and Victor Steinberg}
 \affiliation{Department of Physics of Complex Systems, Weizmann Institute of Science, Rehovot, Israel 76100}

%\date{\today}

\begin{abstract}
It is shown that a channel flow of a dilute polymer solution between two widely spaced cylinders hindering the flow is an important paradigm of an unbounded flow in the case in which the channel wall is located sufficiently far from the cylinders. The quantitative characterization of  instabilities in a creeping viscoelastic channel flow between two widely spaced cylinders reveals two elastically driven transitions, which are associated with the breaking of time-reversal and mirror  symmetries: Hopf and forward bifurcations described by two order parameters $\mbox{v}_{rms}$ and $\bar{\omega}$, respectively. We suggest that a decrease of the normalized distance  between the obstacles leads to a collapse of the two bifurcations into a codimension-2 point, a situation general for many non-equilibrium systems. However, the  striking and unexpected result is the discovery of a  mechanism of the vorticity growth via an increase of a vortex length at the preserved streamline curvature in a viscoelastic flow, which is in  sharp contrast to the well-known suppression of the vorticity in a Newtonian flow by polymer additives.

\end{abstract}

%\pacs{47.20.Gv, 47.50.Ef, 47.50.Gj}

\maketitle

  The addition of a small amount of high-molecular-weight flexible polymer molecules into a fluid strongly affects a flow in a wide range of spatial and temporal scales. Polymers being stretched by a velocity gradient particularly in a flow with curvilinear streamlines generate elastic (hoop) stresses, which react back on the flow and modify it via elastic instabilities and at further stretching lead to elastic turbulence (ET) discovered at $Re\ll1$ \cite{groisman,myreview}. Elastic turbulence is a chaotic flow characterized by a strong enhancement of a flow resistance, a power-law decay of velocity power spectra with an exponent $|\alpha|> 3$, and orders of magnitude enhancement of mixing compared with diffusion \cite{groisman1,groisman2,myreview}.
 Theory \cite{lebedev} and numerical simulations \cite{berti,berti1,yatou,khomami,grilli} of ET consider unbounded, homogeneous, and isotropic flow of a dilute solution of polymers with linear elasticity that is strongly  distinguished from bounded, anisotropic, and inhomogeneous flow studied experimentally \cite{groisman1,groisman2,teo1,teo2,jun,liu, jun2}. There are two approaches to resolve the evident discrepancy: either to look for flow geometry, where unbounded, homogeneous, and isotropic flow of ET can be realized experimentally, or to develop ET theory for a bounded container with a nonzero mean velocity $\bar{\mbox{u}}$.

 In our search for the experimental realization of an unbounded, homogeneous, and isotropic flow of a viscous polymer solution at $Re\ll1$, we consider a flow past an obstacle or array of obstacles. In spite of the fact that the flow past the obstacle is considered as a paradigmatic problem of fluid mechanics for both Newtonian and viscoelastic fluids widely investigated in the past both experimentally and numerically and is highly relevant to many industrial applications, the studies of highly elastic fluids at $Re\ll1$ in such a flow geometry are rather limited.

 In a viscoelastic creeping flow past a cylinder, three nondimensional parameters control the dynamical behavior of the flow, namely, the Weissenberg number $Wi=\lambda\bar{\mbox{u}}/2\mbox{R}$ and two geometrical parameters: the blockage ratio $b$ of the cylinder diameter $2\mbox{R}$ to the channel width $w$, $b=2\mbox{R}/w$, and the confinement ratio $a$ of the channel height $h$ to width, $a=h/w$. Here $Wi$ is the ratio of the nonlinear elastic stress to its dissipation via relaxation and defines  a degree of polymer stretching, $\bar{\mbox{u}}$ is the average flow velocity, and $\lambda$ is the longest polymer relaxation time \cite{bird}. The parameter $b$ controls the relative strength of shear near the wall and extension near a stagnation region, whereas $a$ controls the two-dimensional (2D) versus 3D effects. For both small $a$ and $b$ one expects mainly 2D confined flow near an unbounded cylinder with large extensional strains.

 At $b\ll1$, two approaches have been explored in experiments as well as simulations: to study a friction coefficient $f$ either of a free-falling body (cylinder or sphere) at a terminal velocity or of a channel flow past a cylinder or an array of cylinders. In the former case, a substantial number of experimental studies with various polymer solutions have been conducted and controversial results even on $f$ have been obtained \cite{james}. Reliable experimental data on $f$ are reported in Ref. \cite{muller}; however, the quantitative discrepancy between its value and numerical simulations remains unresolved \cite{khomami1}.

 At moderate $Wi$, a detailed investigation of a viscoelastic creeping flow past a  cylinder reveals a stationary elastic wake instability, which leads to the formation of a 3D regularly spaced cellular structure \cite{McKinley}. A further increase in $Wi$ results in a subsequent transition to a time-dependent flow \cite{McKinley}. Subsequent experiments in a microchannel flow past a strongly confined cylinder with various $a$ and $b$ and in a wide range of $Wi$ and $Re$ were conducted in Ref. \cite{kenney}. In this work, a downstream elastic instability is observed only at $Wi\geq83$ and  $Re\geq 1$, which is not the creeping flow discussed here. Moreover, an upstream instability at higher $Wi$ and $Re$ was also reported \cite{kenney}.

 More extensive studies of a purely elastic instability were conducted both experimentally and numerically in a creeping flow in a wall-bounded channel with a periodic array of cylinders. For closely spaced cylinders, a pair of vortices between cylinders was observed for both Newtonian and viscoelastic fluids at $Wi$ below an instability at $Wi_c\simeq 1.1$ \cite{khomami2,khomami3}. Above $Wi_c$, a sharp increase in $f$ related to the onset of noisy oscillations of a cross-stream velocity  due to breaking of  time-reversal symmetry was found. The coupling between the hoop stress arising from curved streamlines and the velocity perturbations plays a key role in triggering the instability \cite{khomami3}. It resembles an oscillatory instability in an extensional viscoelastic flow realized in T-junction geometry with a long recirculating cavity \cite{atul}. Two-dimensional numerical simulations of the viscoelastic flow for closely spaced cylinders reveal an instability associated with the cross-stream velocity fluctuations at $Wi_c\approx1.1$ resulting from a time-dependent instability of the vortex pairs, generated at $Wi<Wi_c$, in  agreement with the experiments \cite{ellero}. Moreover, both the growth of $f$ and root-mean-square (rms) fluctuations of the cross-stream velocity with $Wi$ at $Wi>Wi_c$ reveals a square-root dependence on $Wi$ characterized as the forward bifurcation \cite{cross}. Three-dimensional simulations in \cite{grilli} reproduce the $f$ growth with $Wi$ at $Wi>Wi_c$, the same as in a 2D flow, pointing out the 2D nature of perturbations causing the instability.

 In this Letter we present experimental results of a viscoelastic creeping channel flow instability between two widely-spaced cylinders, which provide a quantitative answer to the following questions.  How is a velocity field  modified by elasticity in an unbounded ($b\ll1$) flow between two cylinders and what is the corresponding order parameter of the elastic instability? How do polymers alter the flow resistance as a result of the instability? What are the flow velocity field and its spectral properties and what is the flow structure between the cylinders?

\begin{figure}[htbp]
	\begin{center}
		\includegraphics[scale=0.35]{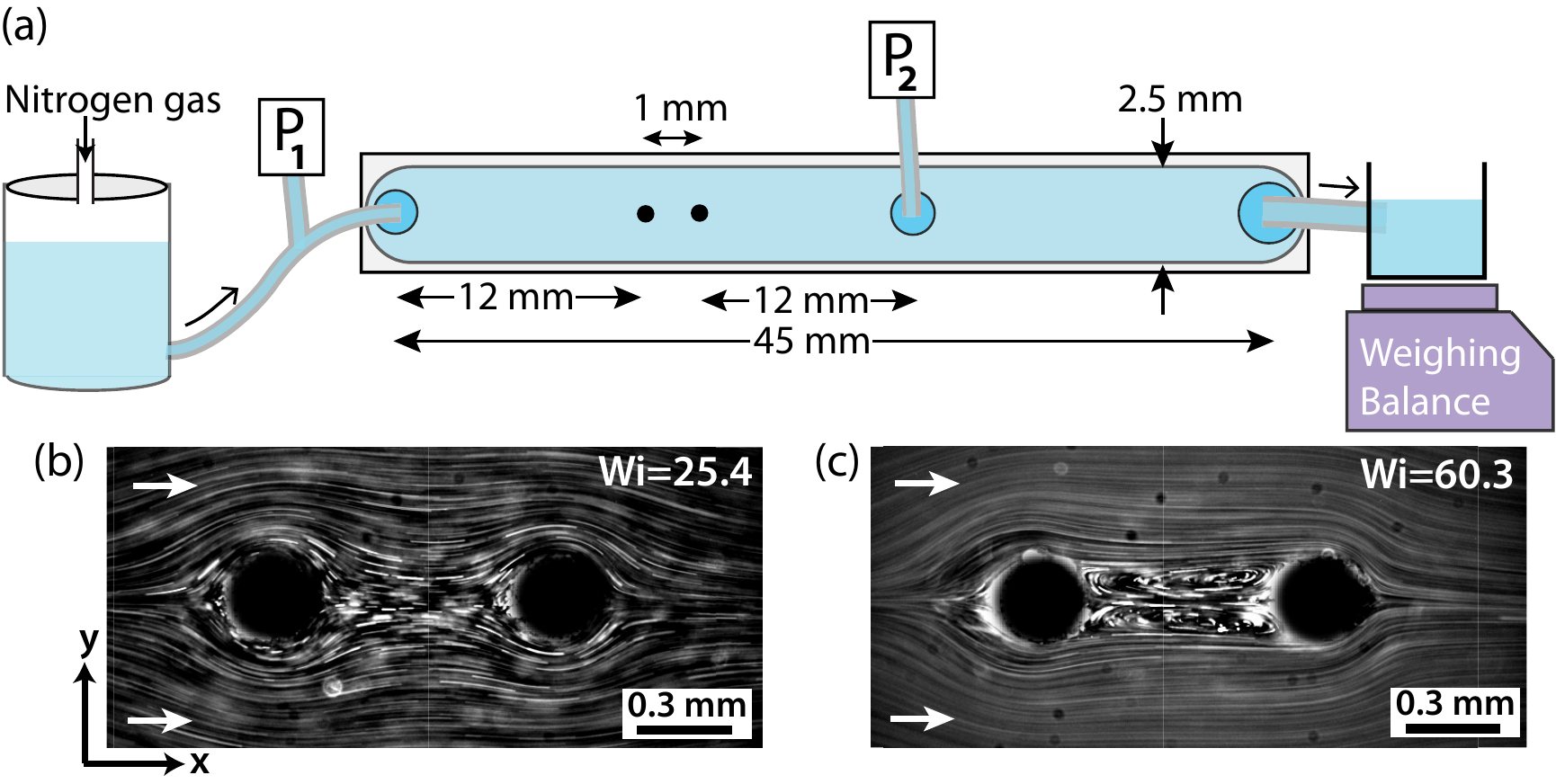}
		\caption{(a) Schematic of experimental setup. The two black dots 1 mm apart are obstacles. Particle streaks of the flow around and between the obstacles at (b) $Wi=25.4$ and $Re=0.002$ and (c) $Wi=60.3$ and $Re=0.0046$. White arrows indicate the flow direction.}
		\label{fig:setup}
	\end{center}
\end{figure}
  A dilute polymer solution of high-molecular-weight polyacrylamide (molecular weight $M_w=18$ MDa, Polysciences) at a concentration $c=100$ ppm ($c/c^*\simeq0.5$, where $c^*=200$ ppm for the polymer used \cite{liu2} is the overlap concentration) is prepared in a viscous solvent of $62\%$ sucrose and $1\%$ NaCl by weight. The solvent viscosity $\eta_s$ at $20^{\circ}\mbox{C}$ is measured to be $0.13~Pa\cdot s$ in a commercial rheometer (AR-1000; TA instruments). The addition of the polymer to the solvent increases the solution viscosity $\eta$ up to $0.15~Pa\cdot s$.  The stress-relaxation method \cite{liu2} is employed to obtain $\lambda=10\pm0.5$ s.  The fluid is driven by nitrogen gas at a pressure up to $\sim5$ psi and is injected via the inlet into a linear channel of dimension $L\times w\times h=45\mbox{mm} \times 2.5\mbox{mm} \times 1\mbox{mm}$, shown schematically in Fig. \ref{fig:setup}(a).  The fluid flow is hindered by the two widely-spaced cylindrical obstacles of $2\mbox{R}=0.30~\mbox{mm}$ made of stainless steel separated by a distance of $e=1~\mbox{mm}$ and embedded at the center of the channel. Thus the geometrical parameters of the channel are $b=0.12$ and $a=0.4$ and the normalized distance between the cylinders $E=e/2\mbox{R}=3.3$.

 Two piezoresistive pressure sensors (ABP series, Honeywell) measure the fluid pressure at two locations: before the channel inlet and after the obstacles, marked with $P_1$ and $P_2$, respectively, in Fig. \ref{fig:setup}(a). The fluid exiting the channel outlet is weighed instantaneously $W(t)$ as a function of time $t$ by a PC-interfaced balance (BA210S, Sartorius) with a sampling rate of $5~\mbox{Hz}$ and a resolution of $0.1~\mbox{mg}$. The time-averaged fluid discharge rate $\bar{Q}$ is estimated as $\overline{\Delta W/\Delta t}$. For flow visualization, the solution is seeded with fluorescent particles of diameter $1~\mu m$ (Fluoro-Max green fluorescent, Thermo Scientific). The region between the obstacles is imaged in the mid-plane directly {\it{via}} a microscope (Olympus IX70), illuminated uniformly with a light-emitting diode (Luxeon Rebel) at $447.5~\mbox{nm}$ wavelength, and a CCD camera (GX1920; Prosilica) attached to the microscope records about 5000 images of resolution $1936\times 1456$ pixels at a rate of $65~fps$. We use particle image velocimetry (PIV) to obtain the spatially resolved velocity  $\vec{\mbox{U}}=(\mbox{u},\mbox{v})$  in the region between the cylinders \cite{piv}.  An interrogation window of $8\times8~pixels^2$ ($24\times24~\mu m^2$) with $50\%$ overlap is chosen to procure $\vec{\mbox{U}}$.

 Figures \ref{fig:setup}(b) and \ref{fig:setup}(c) display two streak flow images at two $Wi$. At lower $Wi=25.4$, the flow between the obstacles is close to a potential one and similar to the potential flow of a Newtonian fluid at $Re<1$, whereas at $Wi=60.3$ the streaks of the inner flow velocity are much shorter than those of an outside flow and vortices are clearly identified (see also movies 1 and 3 in Ref. \cite{sm}). These images qualitatively illustrate the difference of the flow field between the obstacles below and above the instability onset.
\begin{figure}[t!]
	\begin{center}
		\includegraphics[scale=0.25]{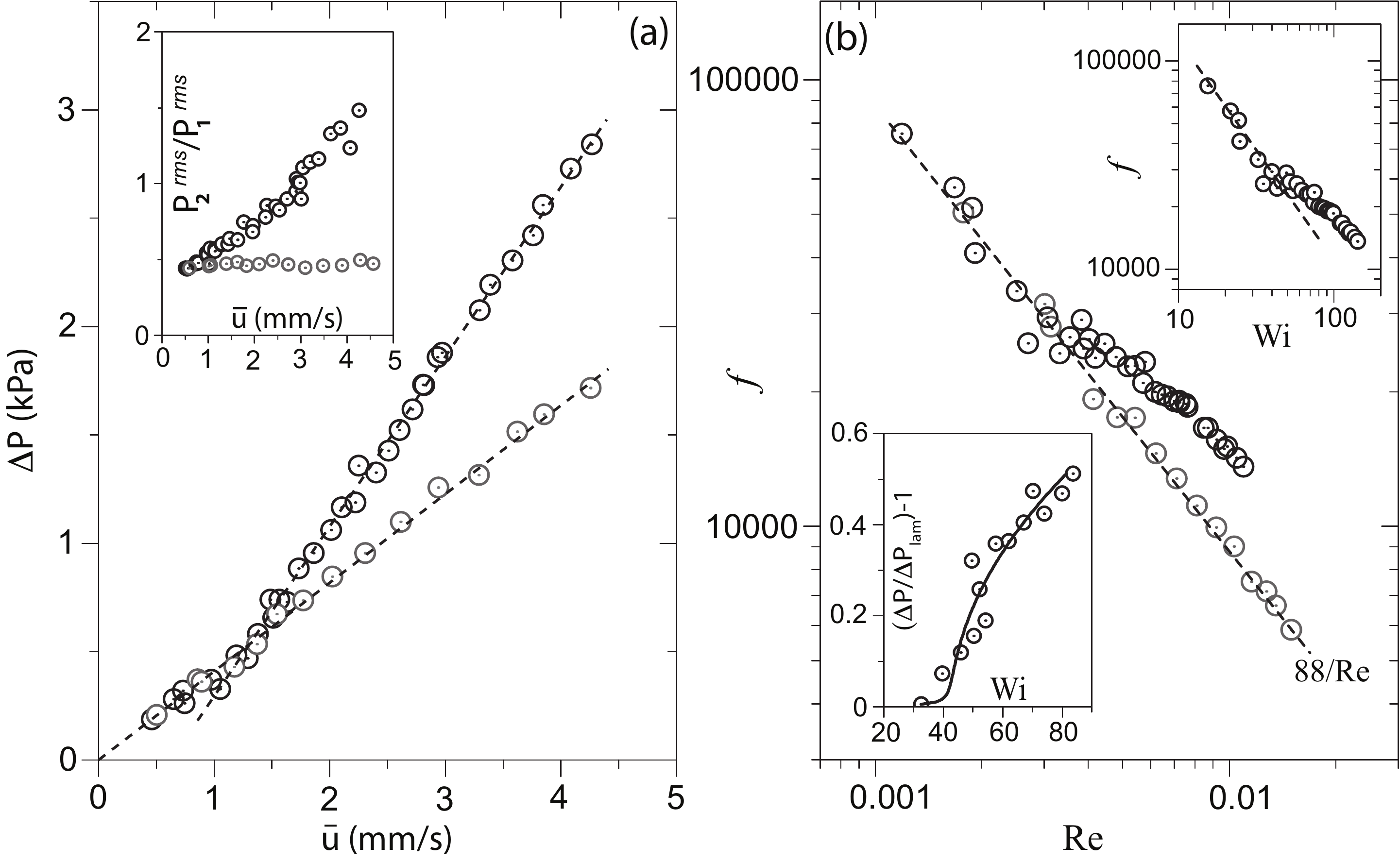}
		\caption{(a) Pressure drop in the channel $\Delta P$ versus $\bar{\mbox{u}}$ for Newtonian (gray circles) and polymer (black circles) fluids.  Inset: $P_2^{rms}/P_1^{rms}$ versus $\bar{\mbox{u}}$. (b) Friction factor $f$ versus $Re$ with the fit for potential flow (dash line). The top inset shows $f$ versus $Wi$. The bottom inset shows $(\Delta P/\Delta P_{lam})-1$ versus $Wi$ with a fit (solid line) (see the text).}
		\label{fig:dragcoeff}
	\end{center}
\end{figure}

 A quantitative characterization of the elastic wake instability through the dependence of the pressure drop   in the channel $\Delta P=P_2-P_1-\Delta P_{pipe}$  on the average flow  speed $\bar{\mbox{u}}$ is presented in Fig. \ref{fig:dragcoeff}(a). The transition is identified by the deviation of $\Delta P$  from a linear dependence of $\Delta P$ on $\bar{\mbox{u}}$ characteristic of a potential flow past  obstacles in a Newtonian solvent and the polymer solution at low $Wi$, as shown in Fig. \ref{fig:dragcoeff}(a) together with the linear fit. Moreover, the rms normalized pressure fluctuations $P_2^{rms}/P_1^{rms}$ grow with $\bar{\mbox{u}}$ for  polymer solution above the transition, as shown in the inset in Fig. \ref{fig:dragcoeff}(a), whereas for the Newtonian solvent  $P_2^{rms}/P_1^{rms}$  remains constant. Here, $\bar{\mbox{u}}=\bar{Q}/\rho wh$ and the pressure drop along the small pipe connected to the inlet of the channel is estimated \cite{moody} from  $\Delta P_{pipe}=(\frac{64}{Re_p})(\frac{\rho L_p \bar{Q}^2}{2\pi^2 r_p^5})$, where $Re_p=(\frac{2r_p\rho}{\eta})(\frac{\bar{Q}}{\pi r_p^2})$ is the Reynolds number for the pipe flow, the fluid density $\rho=1286~kg/m^3$, and $r_p=0.26$ mm and $L_p=72$ mm are  the radius and length of the pipe, respectively. To highlight more clearly the transition region, the  data are shown in Fig. \ref{fig:dragcoeff}(b) on a high-resolution plot via the dependence of the friction factor $f=2D_h\Delta P/\rho \bar{\mbox{u}}^2 L_c$ on the Reynolds number $Re=2\mbox{R}\bar{\mbox{u}}\rho/\eta$, where the
hydraulic radius is $D_h=2wh/(w+h)=1.43$ mm and $L_c=25$ mm is the distance between the inlet and the location of pressure measurement in the channel $P_2$ (marked in Fig. \ref{fig:setup}(a)).
In the main plot the data for the Newtonian solvent in the whole range of $Re$ are described by the fit $f\sim 88/Re$ with a large scatter before the elastic instability. In the top inset in Fig. \ref{fig:dragcoeff}(b), the same data are plotted as $f$ versus $Wi$  and the transition at $Wi_c\approx 40$ is identified as well so the scatter in the vicinity and before $Wi_c$ is rather large.
\begin{figure}[t!]
\begin{center}
\includegraphics[scale=0.33]{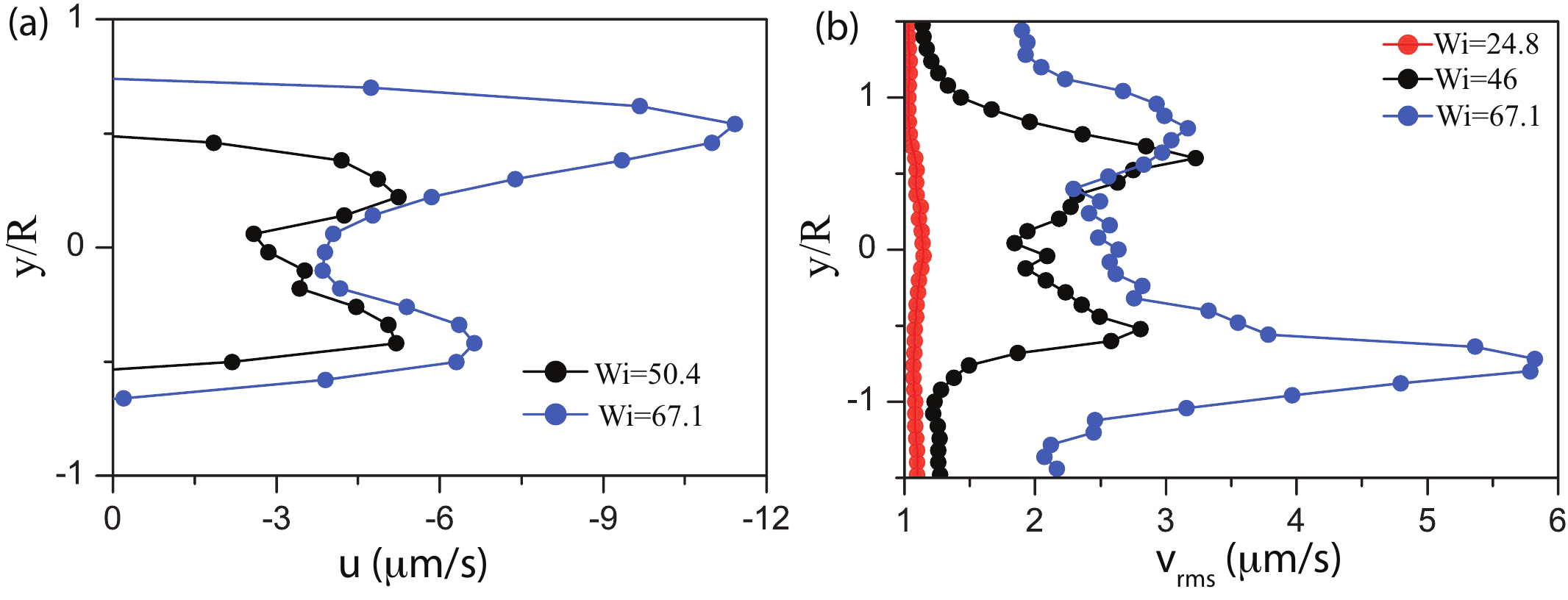}
\caption{(a) Negative $\mbox{u}(\mbox{y}/\mbox{R})$ profiles at $\mbox{x}/\mbox{R}=5.2$ for two $Wi$. (b) Root-mean-square values of cross-stream velocity fluctuations $\mbox{v}_{rms}$ as a function of $\mbox{y}/\mbox{R}$ for three $Wi$ obtained at $\mbox{x}/\mbox{R}=4.2$.}
\label{fig:velocprof}
\end{center}
\end{figure}

 Figure 1SM in Ref. \cite{sm} illustrates  the time-averaged streamwise velocity profile obtained in the horizontal midplane by PIV as a function of $\mbox{y}/\mbox{R}$ for several $\mbox{x}/\mbox{R}$ locations in the region between the obstacles  at four $Wi$ below and above the transition. Here $\mbox{x}$ and $\mbox{y}$ are longitudinal and transverse coordinates of the channel, respectively, with ($\mbox{x},\mbox{y}$)=($0,0$) located at the center of the upstream cylinder. At $Wi<Wi_c$, the flow is in the forward direction everywhere between the obstacles, while at $Wi>Wi_c$ a reverse flow is developed that is demonstrated in the insets by  negative values of $\mbox{u}$ between the obstacles (see Fig. 1SM in Ref. \cite{sm}). A further magnified negative velocity profile is presented in Fig. \ref{fig:velocprof}(a) at $\mbox{x}/\mbox{R}=5.2$ for two $Wi$, where strong velocity gradients  are developed that lead to a significant polymer stretching. It is reminiscent of sharp radial velocity gradients observed in the core of a solitary vortex pair resulting from a pure elastic instability in a viscoelastic Couette flow \cite{groisman3}.
\begin{figure}[t!]
\begin{center}
\includegraphics[scale=0.2]{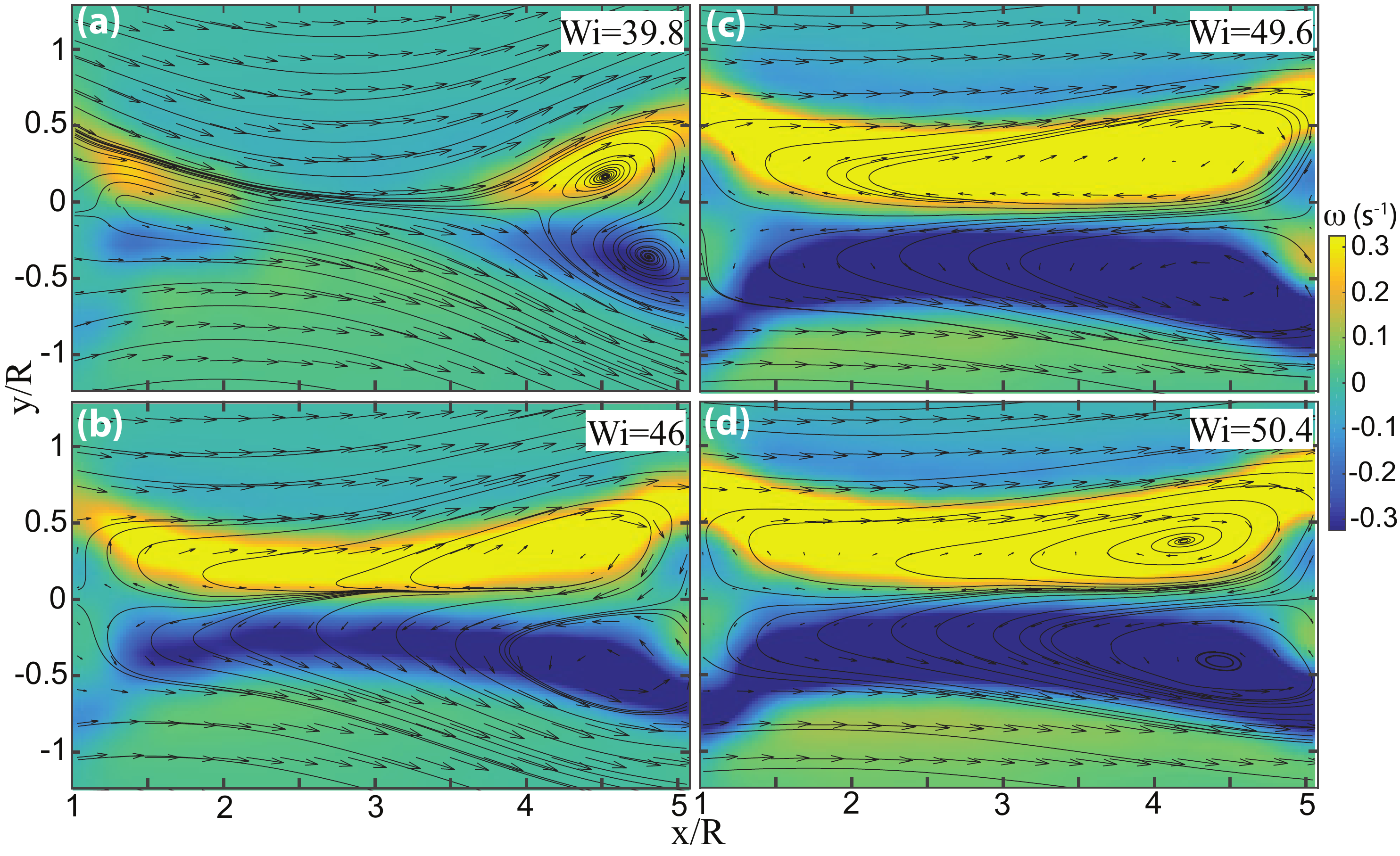}
\caption{Velocity field (shown by arrows) and the resultant vorticity map (in color) obtained by PIV for different $Wi$: (a) $Wi=39.8$, (b) $Wi=46$, (c) $Wi=49.6$, and (d) $Wi=50.4$. Streamlines are shown by the black lines. }
\label{fig:vorticity}
\end{center}
\end{figure}

\begin{figure}[t!]
\begin{center}
\includegraphics[scale=0.33]{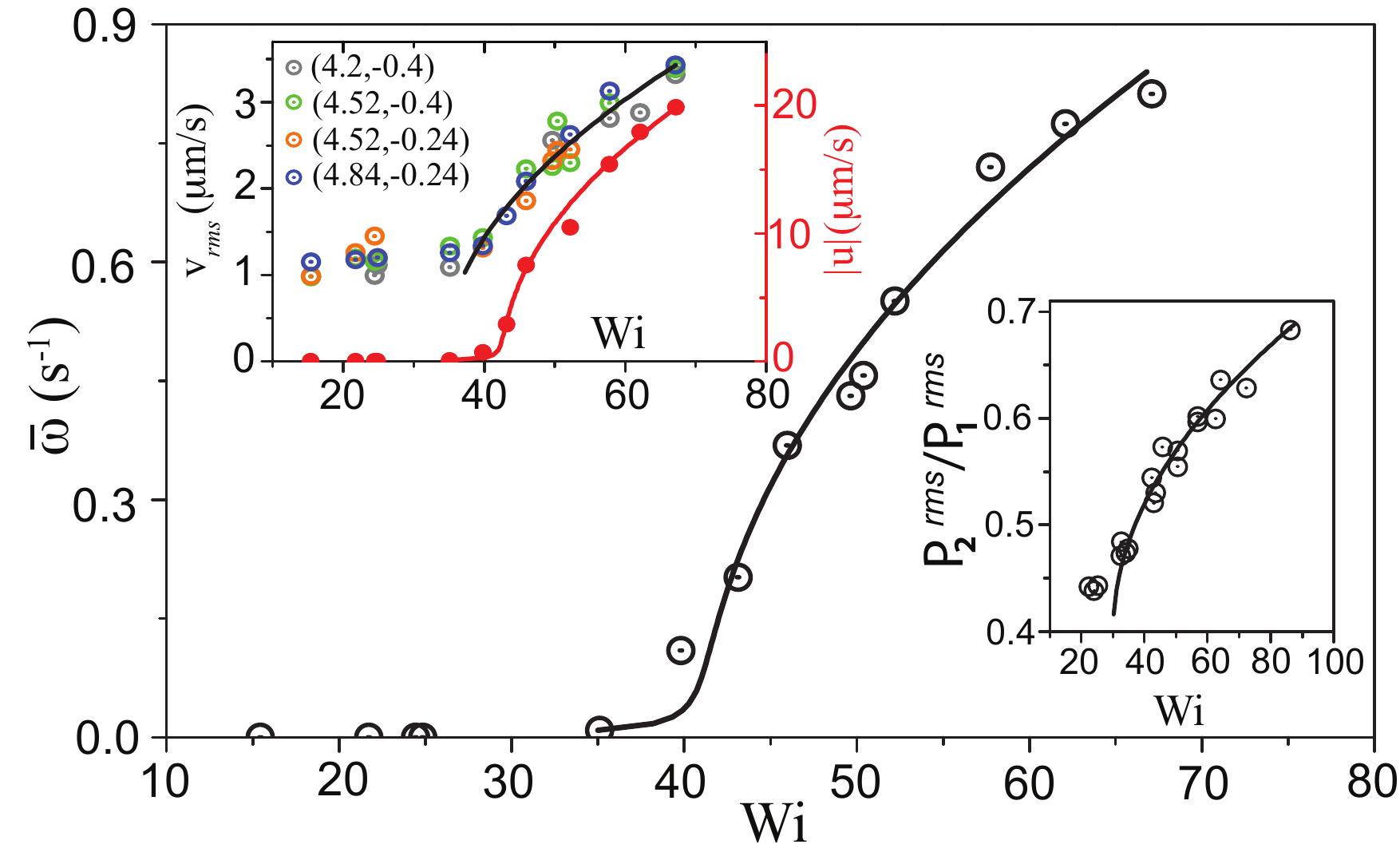}
\caption{Spatially averaged vorticity $\bar{\omega}$ versus $Wi$. Solid line is a fit (see the text). The top inset shows $\mbox{v}_{rms}$ (left ordinate) and the absolute value of the time-averaged reverse flow velocity $\mbox{u}$ (right ordinate) versus $Wi$ at ($\mbox{x}/\mbox{R}, \mbox{y}/\mbox{R}$)=($4.9,-0.4$).  Solid lines are fits (see the text). Bottom inset: $P_2^{rms}/P_1^{rms}$ versus $Wi$ with a fit (solid line) (see the text).}
\label{fig:orderparam}
\end{center}
\end{figure}

 To analyze further the velocity field $\vec{\mbox{U}}$, we compute time-averaged vorticity $\omega$ as $\nabla\times\vec{\mbox{U}}$. Figure \ref{fig:vorticity} shows the vorticity map  between the obstacles for $Wi\ge Wi_c$. A small vortex pair appears first at $Wi=34.1$ (see movie 2 in Ref. \cite{sm}) in the vicinity of the downstream cylinder and expands in size with $Wi$, as shown in Fig.  \ref{fig:vorticity}.
The quantitative dependence of the spatially averaged, either positive or negative, vorticity $\bar{\omega}$ on $Wi$ is presented in  Fig. \ref{fig:orderparam} together with the fit based on the Landau equation for the order parameter of the continuous transition $[(Wi/Wi_c)-1]\bar{\omega}-c\bar{\omega}^3+d=0$ \cite{cross}, which yields $Wi_c=41.6\pm 0.9$ as the fit parameter. A similar fit is used to characterize the transition in the dependence of $(\Delta P/\Delta P_{lam})-1$ on $Wi$, as shown in the bottom inset in Fig. \ref{fig:dragcoeff}(b), where a close value of $Wi_c=43.3\pm 2.4$ but with significantly larger scatter is obtained. Here $\Delta P_{lam}$ is the pressure drop for a potential flow (see Fig. \ref{fig:dragcoeff}(a)). The top inset in Fig. \ref{fig:orderparam} shows the dependences of an absolute value of the reverse streamwise flow velocity $|\mbox{u}|$ at $(\mbox{x}/\mbox{R},\mbox{y}/\mbox{R})=(4.84,-0.4)$ and the rms cross-stream velocity fluctuations $\mbox{v}_{rms}$  at four locations close to the downstream cylinder on $Wi$ that are time averaged for about $70$ s. A similar fit by the Landau equation used for $|\mbox{u}|$ provides $Wi_c=42.8\pm 0.9$, in accord with the value obtained for $\bar{\omega}$, since the vorticity and the reverse flow velocity between the  obstacles are directly related to each other. On the other hand, a fit to the data $\mbox{v}_{rms}$ versus $Wi$ by $\mbox{v}_{rms}\sim(Wi/Wi_c-1)^{0.5}$ provides $Wi_{1c}=34.3\pm0.8$ significantly lower than the values obtained above. It should be emphasized that the $\mbox{v}_{rms}$ values above the transition are an order of magnitude smaller than  $|\mbox{u}|$, which should correspond to their contributions to the friction factor $f$. Indeed, as one can see from the bottom inset in Fig. \ref{fig:dragcoeff}(b), the whole range of change of $(\Delta P/\Delta P_{lam})-1$ is about 0.5. Then a value, about 10 times smaller,  corresponding to the ratio $\mbox{v}_{rms}/|\mbox{u}|$ is inside the scatter and cannot be identified in $f$.  Though from the dependence of $P_2^{rms}/P_1^{rms}$ versus $Wi$ in the bottom inset in Fig. \ref{fig:orderparam} the early transition is clearly characterized by the fit $(Wi/Wi_c-1)^{0.5}$ with $Wi_{1c}\simeq 30\pm2$, close to the above value.

\begin{figure}[t!]
\begin{center}
\includegraphics[scale=0.39]{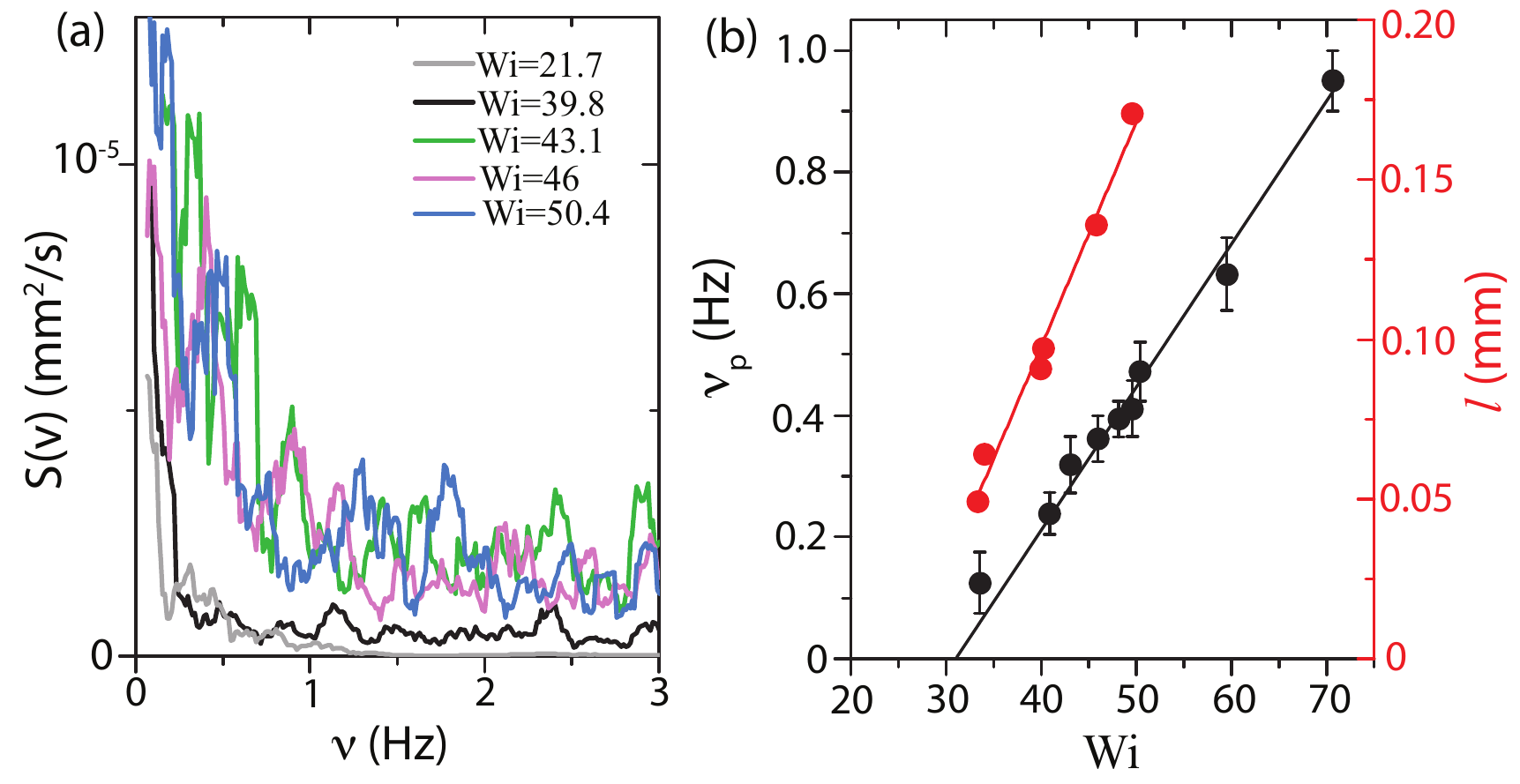}
\caption{(a) Cross-stream velocity power spectra $S(\mbox{v})$ versus frequency $\nu$ at $(\mbox{x}/\mbox{R}, \mbox{y}/\mbox{R})=(4.2,-0.24)$ for five $Wi$. (b) The left ordinate is the dependence of the characteristic frequency $\nu_p$ on $Wi$ above $Wi_{1c}$. The solid line is a fit and its extrapolation to $\nu_p=0$ yields $Wi_{1c}\approx 31$. The right ordinate is the vortex length $\ell$ as a function of $Wi$. The solid red line is a fit. }
\label{fig:spectra}
\end{center}
\end{figure}

 To further understand the nature of the instability associated with $\mbox{v}_{rms}$, we compute the frequency power spectra $S(\mbox{v})$ of the cross-stream velocity $\mbox{v}$ for five $Wi$, as shown on a linear scale in Fig. \ref{fig:spectra}(a). The peaks in $S(\mbox{v})$ are distinctly evident at $Wi>Wi_{1c}$ with a characteristic frequency $\nu_p$ varying linearly with $Wi$ and approaching zero at $Wi_{1c}\approx31$ (see Fig. \ref{fig:spectra}(b)), whereas $S(\mbox{v})$ at $Wi<Wi_{1c}$ is rather smooth. Thus, the first transition at $Wi_{1c}\approx 32$, taken as the average of values obtained from $\mbox{v}_{rms}$ and $P_2^{rms}/P_1^{rms}$, is indicative of the Hopf bifurcation \cite{cross,atul}.
From profiles of $\mbox{v}_{rms}(\mbox{y}/\mbox{R})$ at $\mbox{x}/\mbox{R}=4.2$ for three $Wi$ one finds that at $Wi>Wi_c$ the velocity fluctuations are much higher than an instrumental noise level and two peaks in the profiles indicate the border between the outer and inner velocity flow regions (see Fig. \ref{fig:velocprof}(b)). According to the peak locations the inner region widens with $Wi$. Similar growth of $\mbox{v}_{rms}$ characterizes an elastic instability observed in 2D numerical simulations of a viscoelastic channel flow past a periodic array of closely spaced ($E=1.25$) cylinders, where the velocity power spectra and the $\mbox{v}_{rms}\sim (Wi/Wi_{1c}-1)^{0.5}$ scaling at $Wi>Wi_c$ are presented in Figs. 9 and 10 of Ref. \cite{ellero} and in experiments \cite{khomami2,khomami3}. The second steady forward bifurcation is not observed in this case probably due to the difference in $E=1.25$ versus 3.3 used in our experiment. Thus, in the current experiment two subsequent transitions, close in $Wi$ values, are found: first at $Wi_{1c}\approx 32$ as the Hopf bifurcation and second at $Wi_{2c}\approx 42$ as the forward bifurcation. In the latter case, $Wi_{2c}$ is taken as the average of values obtained from $\bar{\omega}$ and $|\mbox{u}|$. The value of $Wi_{2c}$ found from the dependence of $(\Delta P/\Delta P_{lam})-1$ on $Wi$ is  higher with much larger scatter and a lower resolution in the $Wi_{2c}$ detection, as pointed out above.

\begin{figure}[t!]

\begin{center}
\includegraphics[scale=0.39]{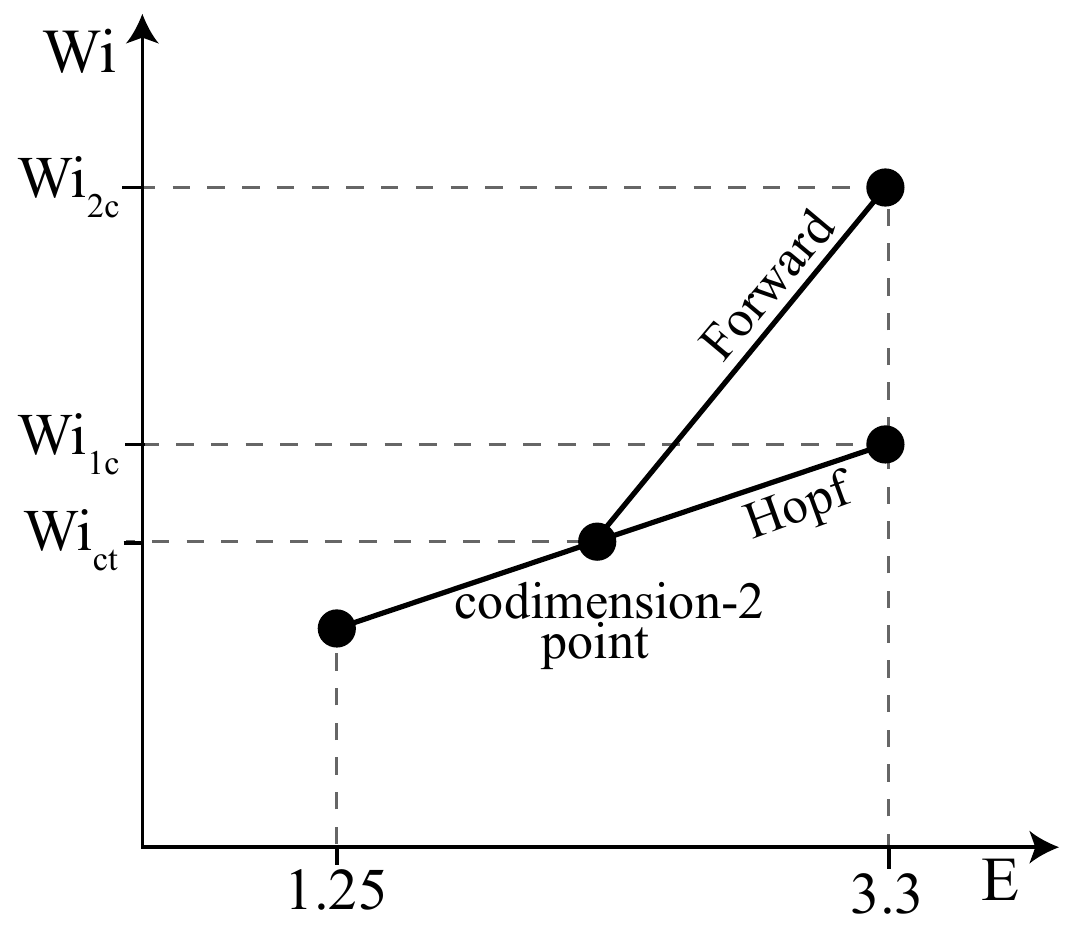}
\caption{Schematic of a phase diagram in the vicinity of codimension-2 point. }
\label{fig:diagram}
\end{center}
\end{figure}

 What is the physics behind the both elastic instabilities described above? A small vortex pair appearing due to a breaking of mirror symmetry generates  the hoop stress $F_h$ due to a curvature, which interacts with the cross-stream velocity perturbations. The latter arises due to the breaking of time-translational invariance. As a result, the Hopf bifurcation appears first at $Wi_{1c}\approx32$ and is similar to that found by us in T-junction geometry with a long recirculating cavity \cite{atul}. A further increase in the external driving leads via the second instability to an enhancement of the vortex vorticity and so the elastic stress $\sigma_{el}$. Indeed, if the vortex preserves its curvature, the hoop stress grows \cite{bird} as $F_h\sim \sigma_{el}/r\sim\bar{\omega}^2/r\sim \mbox{u}^2/r^3$, where $r$ is the vortex radius. Then, above the second transition one gets $F_h\sim \mbox{u}^2\sim (Wi/Wi_{2c})-1$, which finally would cause the vortex collapse due to the growing hoop stress. So this means that the vortex pair would be suppressed. Indeed, for example, an inhibition of a von K\'{a}rm\'{a}n vortex street in a cylinder wake in a Newtonian fluid by an injection of a polymer additive is observed \cite{cadot,goldburg}. The way out from this evident discrepancy is a growing vortex size with $Wi$. If for simplicity we introduce a vortex length $\ell$, then the same estimate leads to $F_h\sim \mbox{u}^2/\ell r^2$. As found from the experiment, $\ell\sim Wi$ at $Wi\geq Wi_{2c}$ (see Fig. \ref{fig:spectra}(b)) and so $F_h\sim \mbox{u}^2/\ell r^2\simeq const$ remains intact and thus the vortex remains stable. The question arises why the second transition was not observed in the previous experiments of a flow past a periodic array of cylinders. The main reason is the difference in the value of $E$  that is larger in the current experiment. So if $E$ decreases, both transitions finally may collapse, leading to a codimension-2 point \cite{steinberg,groisman4,cross}, where two order parameters reach zero simultaneously. A further decrease in $E$ leads to a complete suppression of the forward bifurcation, which is indeed found at $E=1.25$ in both experiments and numerical simulations \cite{grilli,khomami2,khomami3,kenney,ellero} (see Fig. \ref{fig:diagram}).

 To summarize, a channel flow of a dilute polymer solution between two widely spaced cylinders hindering the flow is an important paradigm of an unbounded flow in the case of a channel wall being located sufficiently far from the cylinders. The quantitative analysis of the elastic instabilities in this flow uncovers a rather general sequence of two bifurcations resulting from the breaking of time-reversal symmetry as the first and mirror symmetry as the second, which are associated with two order parameters: $\mbox{v}_{rms}$ and $\bar{\omega}$, respectively. The former experiments and simulations for similar conditions but for closely spaced cylinders have found only the first transition. So $E$ can be considered as the second control parameter, which reducing from 3.3 to 1.25 leads to elimination of the second instability. The latter suggests that the codimension-2 point, where both order parameters approach zero, exists between these two $E$ values. It is a general case in a large variety of nonequilibrium systems \cite{cross}. However, the  striking and unexpected result is the discovery of the mechanism of the vorticity growth due to an increase of the vortex length at the preserved streamline curvature in a viscoelastic fluid flow, which is in  sharp contrast to the well-known suppression of the vorticity in a Newtonian fluid flow by polymer additives \cite{cadot,goldburg}.

 We thank Guy Han and Yuri Burnishev for technical support. This work was partially supported by grants from Israel Science Foundation and Volkswagen Foundation via the Lower Saxony Ministry of Science and Culture Cooperation (Germany).

\end{document}